\documentclass[12pt,preprint]{aastex}

\begin{document}

\title{Autocorrelation function of the soft X-ray background 
       produced by warm-hot gas in dark halos}

\author{Xiang-Ping Wu and Yan-Jie Xue}

\affil{National Astronomical Observatories, Chinese Academy
                 of Sciences, Beijing 100012, China}

\begin{abstract}
We calculate the angular two-point autocorrelation function (ACF) 
of the soft X-ray background (SXRB) produced by the warm-hot 
intergalactic medium (WHIM) associated with dark halos, 
motivated primarily by searching for missing baryons and 
distinguishing different physical processes of the WHIM 
in dark halos.  We employ a purely analytic model
for the halo population which is completely determined by the universal
density profile and the Press-Schechter mass function. We then adopt 
a phenomenological approach to nongravitational processes 
of the WHIM such as preheating and radiative cooling.
It shows that the power spectra of the SXRB predicted by 
three WHIM models, namely, the self-similar model, preheating model 
and cooling model demonstrate remarkably different signatures 
in both amplitude and shape, with the peak locations moving from 
$\ell\approx4\times10^4$ for the self-similar model to a smaller 
value of $\ell\approx(3$-$5)\times10^{3}$  when nongravitational 
processes are taken into account. The corresponding ACFs 
for preheating and cooling models become shallower too as 
compared with the prediction of the self-similar model. This may 
permit an effective probe of the physical processes of the WHIM 
in massive halos in conjunction with the observationally determined  
power spectrum or ACF of the SXRB from diffuse WHIM. 
However, a direct comparison of our theoretical predictions 
with existing data (e.g. the ACF determined from ROSAT observations) 
is still difficult because of the dominant contribution of AGNs in 
the soft X-ray sky. We discuss briefly the implication of
our results for resolving the missing baryon problem 
in the local universe. 
\end{abstract}


\section{Introduction}

Hydrodynamical simulations of structure formation suggest 
that  most of baryons in the local universe may
exist in the form of diffuse warm-hot intergalactic medium (WHIM) 
at temperatures of $T\sim10^5$--$10^7$ K (Cen \& Ostriker 1999).
The WHIM is believed to be associated with groups and poor 
clusters that are further embedded in large-scale structures.
As a result, the gravitationally heated WHIM may manifest 
itself as large-scale X-ray
filamentary structures through thermal bremsstrahlung emission
(Scharf et al. 2000; Zappacosta et al. 2002). However, the majority
of the baryons may escape direct detection due to the sensitivity
limit of current X-ray instruments. A newly proposed approach is  
to search for the resonant absorption from the WHIM in large-scale
filamentary structures (Cen et al. 2001). 
It seems that strong evidence for the 
detection of absorption lines associated with the local WHIM has 
already been discovered (Fang et al. 2002; Nicastro et al. 2002a,b).

It has been realized in recent years that the WHIM 
may also contribute a non-negligible fraction of the cosmic soft X-ray
background (SXRB) (e.g. Pen 1999; Dav\'e et al. 2001; 
Voit \& Bryan 2001a; 
Bryan \& Voit 2001; Croft et al. 2001; Wu \& Xue 2001) and therefore, 
measurements of the strength and power spectrum of the SXRB 
may allow us to set stringent constraints on both the amount and 
distribution of the WHIM in the local universe.
In fact, a tight limit on the SXRB from diffuse gas,
after discrete sources are removed, has been obtained with advanced 
X-ray detectors such as ROSAT, Chandra and XMM 
(e.g. McHardy et al. 1998; Hasinger et al. 1998, 2001;
Mushotzky et al. 2000; Giacconi et al. 2001, 2002; Tozzi et al. 2001;
Hornschemeier et al. 2001; Rosati et al. 2002; Bauer et al. 2002; etc.).
In particular, the angular two-point autocorrelation (ACF) 
and harmonic power spectrum of the SXRB have been measured 
over a broad angular scale ranging from ten degrees to 
a few arcminutes (Barcons et al. 1989; Carrera et al. 1991; 
Soltan 1991; Soltan \& Hasinger 1994; Soltan et al. 1996, 1999;  
Carrera, Fabian, \& Barcons 1997;
Soltan, Freyberg \& Tr\"umper 2001; \'Sliwa, Soltan \& Freyberg
2001; Soltan,  Freyberg \& Hasinger 2002), in which the
signature of the diffuse WHIM has been clearly detected 
at arcminute scales (Soltan et al. 2001).
This is actually comparable with the angular resolution achieved 
very recently by 
cosmic microwave background (CMB) anisotropy measurements, in which 
the excess power relative to primordial anisotropy at arcminute scales
is interpreted as the scattering of CMB photons by the hot electrons 
within clusters, namely, the Sunyaev-Zel'dovich (SZ) effect 
(Dawson et al. 2001; Bond et al. 2002).  
In a sense, the power spectra of both SXRB and CMB at arcminute scales 
provide equally important information on the amount and 
distribution of the WHIM in the universe, although X-ray and SZ signals 
have very different connections with the WHIM. For example, the diffuse 
X-ray emission is sensitive to the clumping structures of 
the WHIM and is also subject to the WHIM enrichment. 
Moreover, the unresolved SXRB after discrete sources are removed  
is primarily produced by the WHIM in the local universe.  By contrast, 
SZ signals are rather insensitive to the location of  
the WHIM. At this point, the SXRB may be more suited for the study of
the missing baryons. 

The theoretical framework of evaluating the ACF of the SXRB was 
established many years ago by Barcons \& Fabian (1988). 
Similar technique  was simultaneously 
developed by Cole \& Kaiser (1988) for the computation of 
the SZ power spectrum. Furthermore, an extensive investigation of 
the harmonic power spectrum of the SXRB has been recently
made by \'Sliwa et al. (2001).  Essentially, the ACF or 
power spectrum of the SXRB can be effectively predicted and 
compared with observations, provided that the distribution and 
evolution of the WHIM as well as the cosmological background
are specified. However, major uncertainties in such an exercise 
arise from our poor knowledge of nongravitational 
processes of the WHIM if we accept the prevailing cosmological model 
(e.g. the concordance model) and hierarchical theory of 
structure formation (e.g. the $\Lambda$CDM model).
This is because the WHIM is easily disturbed by many complicated physical 
processes in addition to gravity and thermal pressure, which include
(non)gravitational heating, radiative cooling, star formation 
and energy feedback, magnetic fields, etc. 
The reliability of our theoretical predictions of the intensity and 
ACF of the SXRB is closely connected with the question of 
how well one can handle these non-gravitational mechanisms.
Based on the analysis of the SXRB intensity alone, it has been shown that
the SXRB produced by the gravitationally heated and 
bound WHIM in groups and clusters within the standard framework of
hierarchical formation of structures vastly exceeds the upper limits 
set by current X-ray observations, arguing for the presence of 
nongravitational heating process during the evolution of the WHIM
(e.g. Pen 1999; Wu, Fabian \& Nulsen 2001; Bryan \& Voit 2001; 
Wu \& Xue 2001). 
On the other hand, it may become possible to use the unresolved SXRB 
as a probe of the underlying physical processes of the WHIM 
(Xue \& Wu 2003). Yet, there is still no agreement 
about the significance of  nongravitational heating 
on the SXRB produced by the WHIM (e.g. Croft et al. 2001; 
Dav\'e et al. 2001; Phillips, Ostriker \& Cen 2001).
This partially reflects the difficulty in handling the 
complex processes of the WHIM evolution.

In this study, we will focus on the nongravitational effect 
of the WHIM on the evaluation of the power spectrum and ACF of the SXRB.
Using hydrodynamical simulations without inclusion of nongravitational 
heating, Croft et al. (2001) attempted to construct the ACF of the SXRB.
Their result is in quantitative agreement with the extrapolated ACF of 
the SXRB measured at scales of  $>20$ arcminutes by Soltan et al. (1999),
which is, however, dominated by AGNs.
By contrast, a recent investigation of Zhang \& Pen (2002), based purely 
on analytical models, has shown that the ACF of the SXRB at arcminute
scales can be significantly modified by the presence of nongravitational 
heating process. Further investigation is thus needed to 
clarify the situation.  Recall that there is growing observational 
evidence for the existence of nongravitational process in groups and
clusters, which includes the significant departure of 
the observed X-ray luminosity - temperature relation of groups 
and clusters from the prediction of the self-similar model 
(e.g. Edge \& Stewart 1991; David et al. 1993; Wu, Xue \& Fang 1999; 
Helsdon \& Ponman 2000; Xue \& Wu 2000 and references therein)
and the entropy excess in the central cores of 
groups and clusters (Ponman, Cannon \& Navarro 1999; Lloyd-Davies,
Ponman \& Cannon 2000).

Two prevailing nongravitational scenarios will be considered in 
the present investigation:  preheating and radiative cooling.
Both of them tend to suppress the X-ray emission of the WHIM heated 
by purely gravitational shocks and compression, and become 
indistinguishable at present in the explanation of the observed 
X-ray properties of groups and clusters (Voit \& Bryan 2001b; 
Voit et al. 2002; Borgani et al. 2002; Xue \& Wu 2003).
Nonetheless, it has been argued that radiative cooling alone 
is probably insufficient to account for the unresolved SXRB  
(Wu et al. 2001; Xue \& Wu 2003), while preheating
model or a combined model of cooling and heating may be able to produce
an SXRB intensity below the observational limit 
(Wu et al. 2001; Voit \& Bryan 2001b; Xue \& Wu 2003).
In the present study we will demonstrate how the ACF of the SXRB is
affected by preheating and radiative cooling processes. 
To achieve this, 
we will employ a simple analytic model to determine the new 
distribution of the WHIM in dark halos when preheating
or radiative cooling is added, as was done by our recent work 
(Xue \& Wu 2003). Incorporating our WHIM models with the distribution 
and cosmic evolution of dark halo model described by the universal 
density profile and the Press-Schechter (PS) (Press \& Schechter 1974)  
formalism, we will be able to work out the power spectrum and ACF of 
the SXRB and discuss the feasibility of measurement.
Throughout this paper we assume a flat cosmological  model 
($\Lambda$CDM) of $\Omega_{\rm M}=0.35$, $\Omega_{\Lambda}=0.65$ 
and $h=0.65$.

\section{Halo approach to the ACF of the SXRB} 

In terms of thermal bremsstrahlung,  
the X-ray surface brightness of a halo of mass $M$ at redshift $z$ is 
\begin{equation}
I(z,M,{\mbox{\boldmath $\theta$}})=
        \frac{1}{4\pi(1+z)^4}\int \epsilon(R,Z)d\chi,
\end{equation}
where $\epsilon$ is the emissivity that is calculated using 
the Raymond-Smith (1977) code at the soft X-ray energy band 0.5-2.0 keV 
for an evolving metallicity model of $Z=0.3Z_{\odot}(t/t_0)$, $t_0$ is 
the present age of the universe, $R$ is the radius from the halo center 
and $R=\sqrt{D_{\rm A}^2\theta^2+\chi^2}$, $D_{\rm A}$ is the angular 
diameter distance to the halo, and the integral is performed along
the line-of-sight $\chi$. The Fourier transform of the X-ray surface 
brightness distribution $I(z,M,{\mbox{\boldmath $\theta$}})$ 
in the case of spherical symmetry for the distribution of the WHIM 
in a halo is
\begin{equation}
I_{\ell}(z,M)=2\pi\int I(z,M,\theta)J_0(\ell\theta)\theta d\theta,
\end{equation}
where $J_0$ is the cylindrical Bessel function.
The distribution and evolution of halo population are described by 
the PS theory:
\begin{equation}
\frac{d^2N}{dVdM}=\sqrt{\frac{2}{\pi}}
    \frac{\bar{\rho}}{M}
    \frac{\delta_{\rm c}}{\sigma^2}   \frac{d\sigma}{dM}
    \exp\left(-\frac{\delta_{\rm c}^2}{2\sigma^2}\right),
\end{equation}
where $\bar{\rho}$ is the mean cosmic density, $\delta_{\rm c}$ is the
linear over-density of perturbations that collapsed and virialized at
redshift $z$, $\sigma$ is the linear theory variance of the mass density
fluctuation in sphere of mass $M=4\pi\bar{\rho}R^3/3$.
We parameterize the power
spectrum of fluctuation $P(k)\propto k^nT^2(k)$ and take the fit
given by Bardeen et al. (1986) for the transfer function of
adiabatic CDM model $T(k)$. The primordial power spectrum is assumed
to be the Harrison-Zel'dovich case $n=1$. The mass variance for
a given $P(k)$ is simply
\begin{equation}
\sigma^2(M)=\frac{1}{2\pi^2}\int_0^{\infty}k^2P(k)W^2(kR)dk,
\end{equation}
in which $W(x)=3(\sin x-x\cos x)/x^3$ is the Fourier
representation of the window function. The amplitude in the power
spectrum is determined using the rms fluctuation
on an $8$ $h^{-1}$ Mpc scale, $\sigma_8$, which will be assumed to be 
$\sigma_8=0.9$.

The power spectrum of the SXRB can be separated into the Poisson
term, which arises from the correlation within a single halo
\begin{equation}
P^{\rm p}(\ell)=\int dz\frac{dV}{d\Omega dz}\int dM
                \frac{d^2N}{dMdV}[I_{\ell}(z,M)]^2,
\end{equation}
and the clustering term, which arises from the correlation between 
two different halos
\begin{equation}
P^{\rm c}(\ell)=\int dz\frac{dV}{d\Omega dz} P^m(\ell/\overline{r},z) 
   \left[\int dM\frac{d^2N}{dMdV}b(M,z)I_{\ell}(z,M)\right]^2,
\end{equation}
where we have introduced the halo-halo biasing $b(M,z)$ relative 
to linear matter power spectrum such that the power spectrum of 
halo $M_1$ and halo $M_2$ can be approximated by 
$P^{\rm hh}(z,M_1,M_2,k)=b(z,M_1)b(z,M_2)P^{\rm m}(z,k)$, 
the small-angle approximation has also been assumed so that 
$k\approx\ell/\overline{r}$, $\overline{r}$ is the comoving distance, 
and $dV/d\Omega$ denotes the comoving volume per 
steradian. We calculate $b(z,M)$ in terms of
the prescription of Mo \& White (1996). Finally, the 
angular two-point ACF of the SXRB is simply
\begin{equation}
\omega(\theta)
       = \frac{1}{2\pi \overline{I}^2} \int_0^{\infty}
         \left[P^{\rm p}(\ell)+P^{\rm c}(\ell)\right]J_0(\ell\theta)\ell d\ell.
\end{equation}
The mean SXRB intensity $\overline{I}$ is given by 
\begin{equation}
\overline{I}=\int dz\frac{dV}{d\Omega dz}\int dM
                \frac{d^2N}{dMdV}
               \left[\frac{L_{\rm X}(z,M)}{4\pi D_{\rm L}^2(z)}\right],
\end{equation}
where $L_{\rm X}$ is the total soft X-ray luminosity of a halo of mass $M$
at redshift $z$, and $D_{\rm L}$ is the luminosity distance.

\section{WHIM models}

A set of analytic models for the WHIM distribution in dark halos have 
been derived by Xue \& Wu (2003). Here we choose three models 
corresponding to three typical physical processes of the WHIM: 
(1)Self-similar model: The WHIM simply follows the dark matter distribution, 
in which no radiative cooling and no extra heating are added to the WHIM; 
(2)Preheating model: An extra, constant energy budget is added to the WHIM 
for all halos; And (3)cooling model: Radiative cooling of the WHIM 
is included. We briefly summarize the properties of the three models below.

\subsection{Dark halos}

We begin with the underlying dark matter profile in a halo of mass $M$ 
at redshift $z$. We adopt the universal density profile suggested by
numerical simulations (Navarro, Frenk \& White 1997; NFW)
\begin{equation}
\rho_{\rm DM}(r)=\frac{\delta_{\rm ch}\rho_{\rm crit}}
                 {(r/r_{\rm s})(1+r/r_{\rm s})^2},              
\end{equation}
where $\delta_{\rm ch}$ and $r_{\rm s}$ are the
characteristic density and length of the halo, respectively, which can be
fixed through the so-called concentration parameter 
$c=r_{\rm vir}/r_{\rm s}$ using 
the empirical fitting formula found by numerical simulations 
(Bullock et al. 2001) 
\begin{equation}
c=\frac{10}{1+z}\left(\frac{M}
  {2.1\times 10^{13}M_{\odot}}\right)^{-0.14}.
\end{equation}       
The influence of the $c$ scatter on the ACF of the SXRB will not be 
considered in this study, which roughly reflects the effect of 
different formation time of dark halos. The virial mass $M$ is defined as 
\begin{equation}
M=\frac{4}{3}\pi r^3_{\rm vir}\Delta_{\rm c}\rho_{\rm crit},
\end{equation}
where  $\Delta_{\rm c}$ denotes the overdensity parameter and 
for a flat, $\Lambda$CDM cosmological model, 
$\Delta_{\rm c}=18\pi^2+82[\Omega_{\rm M}(z)-1]-39[\Omega_{\rm M}(z)-1]^2$,
$\Omega_{\rm M}(z)=\Omega_{M}(1+z)^3/E^2$ and
$E^2=\Omega_{\rm M}(1+z)^3+\Omega_{\Lambda}$. 
Finally, we specify the virial temperature of a dark halo 
in terms of the cosmic virial theorem (Bryan and Norman 1998):
\begin{equation}
kT=1.39\;{\rm keV}\;f_{\rm T}
            \left(\frac{M}{10^{15}\;M_{\odot}}\right)^{2/3}
            \left(h^2 E^2\Delta_{\rm c}\right)^{1/3}.  
\end{equation}
We will take the normalization factor to be $f_{\rm T}=0.8$ below.

\subsection{Self-similar model}

We assume that the WHIM particles in halos simply follow the dark matter 
distribution:
\begin{equation}
n_{\rm e}(r)=\frac{f_{\rm b}}{\mu_{\rm e} m_{\rm p}}\rho_{\rm DM}(r),
\end{equation}
where $n_{\rm e}$ is the number density of electrons, 
$f_{\rm b}$ is the universal baryon fraction, and $\mu_{\rm e}=1.13$
is the mean electron weight. We further assume that the WHIM is isothermal 
with temperature $T=T_{\rm vir}$. Consequently, the X-ray surface 
brightness profile $I(z,M,\theta)$ can be straightforwardly obtained. 
To avoid the divergence of $I(z,M,\theta)$ towards the centers of 
the halos, the NFW profile is replaced by a flat core for the 
central regions of radius $0.01r_{\rm vir}$.
Adopting a temperature profile given by the equation of hydrostatic 
equilibrium alters the result only slightly.  
It is well known that this model yields a significant overestimate of 
the SXRB intensity. Here we use this model as a reference point.

\subsection{Preheating model}

We employ a phenomenological approach to constructing the 
preheating model by simply raising the entropy $S^0(r)$ of the WHIM 
in the self-similar model to a certain level regardless of 
whatever the energy sources would be:
\begin{equation}
S(r)=\Delta S+S^0(r).
\end{equation}
We fix the constant entropy floor $\Delta S$ 
using the observed X-ray luminosity and entropy distributions of 
groups and clusters. It turns out that a constant floor of  
$\Delta S=120$ keV cm$^2$ reproduces nicely both the X-ray luminosity - 
temperature relation and central entropy measurement of groups and 
clusters (Xue \& Wu 2003). 
The new distribution of the WHIM with preheating can be 
obtained by combining the equation of hydrostatic equilibrium 
\begin{equation}
\frac{1}{\mu m_{\rm p} n_{\rm e}(r)}\frac{d[n_{\rm e}(r)kT(r)]}{dr}=
  -\frac{GM_{\rm DM}(r)}{r^2}
\end{equation}   
and the new entropy profile $S(r)=kT(r)/n_{\rm e}^{2/3}(r)$. The latter 
acts as the equation of state for the preheated WHIM.

\subsection{Cooling model}

Radiative cooling is completely governed by the conservation of energy:
\begin{equation}
\frac{3}{2}n_{\rm t} kT=\epsilon(n_{\rm e},T,Z)t_{\rm c},
\end{equation}
where $n_{\rm t}$ is the total number density of the WHIM.
Setting the cooling time $t_{\rm c}$ to equal the cosmic age
at the halo redshift  gives the cooling radius 
$r_{\rm cool}$ and total mass $M_{\rm cool}$ of the cooled material. 
Following the prescription of Voit \& Bryan (2001b) and Wu \& Xue (2002a), 
we can find the distribution of the remaining WHIM after cooling 
by solving the equation of hydrostatic equilibrium under the conservation of 
total baryonic mass and entropy. Two uncertainties in this model 
arise from the metallicity $Z$ and the cooling time $t_{\rm c}$. Note that
$t_{\rm c}$ has a cosmological dependence if $t_{\rm c}$ is set to equal the 
age of the universe.

\section{Results}

Figure 1a - 1c show the ACFs of the SXRB over angular scales from
$0.01^{\circ}$ to $1^{\circ}$ predicted by our self-similar
model, preheating model and cooling model, respectively.
A glimpse of these plots reveals that 
overall features of the resulting ACFs look quite similar for all 
three models: The ACFs of the SXRB are dominated by 
the Poisson distribution of the WHIM halos, and the contribution of
halo clustering is essentially negligible. Moreover,
the ACFs of the SXRB are governed by nearby halos within
redshift $z\approx0.2$. In other words, if the local population 
of groups and poor clusters  
can be properly removed in the measurement of the SXRB, the ACF amplitude 
will be significantly reduced especially at large angular separations 
$\theta>0.1^{\circ}$ where the halo clustering becomes the 
dominant component.  Yet,  there is a remarkable difference
between the ACF predicted by the self-similar model and those by
preheating and cooling models. Namely, the ACF power 
in the self-similar model arises primarily from poor groups in 
mass range between $10^{13}$ and $10^{14}$ M$_{\odot}$, while
more massive groups and poor clusters ($10^{14}$-$10^{15}$ M$_{\odot}$)
play a prominent role in the ACFs of the SXRB for preheating and cooling 
models. This is because either the WHIM cannot be trapped in low-mass halos 
in the scenario of preheating or most of the WHIM in low-mass halos
was converted into the cooled materials in radiative cooling model. 

We compare in Figure 2 the ACFs of the SXRB predicted by 
our three WHIM models. 
In addition to their difference in amplitude, the ACFs in preheating and 
cooling models become shallower in shape than the ACF 
in the self-similar model. This arises from the fact that 
both preheating and radiative cooling tend to flatten the
radial profiles of the WHIM inside dark halos, and the effect is more 
remarkable for low-mass halos (e.g. Balogh, Babul \& Patton 1999; 
Babul et al. 2002; Wu \& Xue 2002a,b; Voit et al. 2002), 
which is consistent with the result of Zhang \& Pen (2002) that
the more an extra energy is added to the WHIM, the shallower 
the ACF of the SXRB would be. Yet, 
the most efficient way of distinguishing different WHIM models
is perhaps to work directly with the power spectrum of the SXRB produced 
by the WHIM.  To see this, we have also illustrated in Fig. 2 the  
SXRB power spectra predicted by the three WHIM models. Indeed, 
the inclusion of nongravitational processes (preheating or cooling)
results in a dramatic change in the SXRB power spectrum, with the 
peak location moving from $3\times10^4$ for self-similar model
to $\sim3$-$5\times10^3$ for preheating and cooling scenarios. 
Apparently, this is because X-ray emission is sensitive to 
the dense WHIM in the central cores of dark halos, 
while both preheating and cooling yield a flattening of 
the WHIM distribution in halos. 
As a result, the SXRB power spectrum is peaked 
toward smaller $\ell$ in the presence of nongravitational 
processes as compared with the prediction of the self-similar model.   
Future measurement of the SXRB power spectrum by  
diffuse WHIM at 1-10 arcminute scales should allow us 
to place stringent constraints on the spatial distribution and 
physical processes of the WHIM. Recall that the SZ power spectrum
at $\ell<10^4$ is rather 
insensitive to the physical processes of the WHIM.

We now compare our predictions with the existing results of X-ray 
observations, hydrodynamical simulations and other theoretical work. 
Figure 3 shows our model predictions with and without
`local' population within $z=0.2$, along with the observationally 
determined ACFs of the SXRB by Soltan et al. (2001) and the 
numerically simulated result by Croft et al. (2001) 
without inclusion of preheating. However, 
a direct comparison of our theoretically expected ACFs 
with these observational and simulated data is inappropriate 
because the latter have included contributions from both
point sources like AGNs and extended sources such as galaxies, 
groups and clusters. A reasonable comparison can be made
only if the major component of the SXRB, i.e. AGNs, is properly removed.
In addition, the ACF constructed by Soltan et al. (2001) has excluded 
the pointings centered at local extended sources. This also  
complicates our comparison with theoretical predictions because 
the inclusion of nearby clusters can change 
the predicted ACF substantially. In a word, the current measurements of
the SXRB ACF still do not facilitate a meaningful comparison with
theoretical expectation.   
For a preheating model, we can indeed compare our prediction with 
a similar theoretical investigation by Zhang \& Pen (2002). It 
appears that the two results are roughly consistent with each other 
except at very large angular separations.

\section{Discussion and conclusions}

Using a halo approach to the abundance and evolution of
dark halos and a phenomenological approach to including 
nongravitational heating and radiative cooling for the WHIM,
we have calculated the power spectra and ACFs of the SXRB
produced by diffuse WHIM associated with dark halos.  
Because both nongravitational heating and radiative cooling
tend to produce a flattening of the WHIM distribution in halos,  
the resulting ACFs of the SXRB thus become shallower in shape as 
compared with the one given by the self-similar model. In particular,
the corresponding SXRB power spectra exhibit a signature 
that differs dramatically from the one in the self-similar model, 
with the peak location moving from $\sim3\times10^4$ for
the self-similar model to $(3$-$5)\times10^3$ for 
preheating and cooling models. Therefore,  measurement of
the SXRB power spectrum can be used as an effective tool for
the probe of different physical processes of the WHIM  
at various scales. Unfortunately, the current measurement based on
ROSAT observations is still unable to facilitate such 
an investigation because the available ACF of the SXRB is primarily 
produced by AGNs (Soltan et al. 2001). One possible way
is to subtract the AGN component from the SXRB power spectrum 
using the 3D spatial correlation function of the AGNs that
can be constructed separately. Yet, this may also introduce 
large uncertainties.

One of our initial goals in the present investigation  
is to study the feasibility of searching for the missing
baryons from measurement of the power spectra and 
ACFs of the SXRB. If the missing baryons are the WHIM 
associated with groups and poor clusters that are further
embedded in large-scale structures as shown by hydrodynamical
simulations (Cen \& Ostriker 1999; Dav\'e et al. 2001),
a comparison of the theoretically predicted 
power spectrum, ACF or total intensity of the SXRB and 
the X-ray observed quantities should allow us to set useful constraints on
the content and distribution of the missing baryons. 
Indeed, in the more realistic case of where nongravitational 
effect is taken into account, the major contribution to
the SXRB comes from local groups and poor clusters with
masses ranging from $10^{14}$ to $10^{15}M_{\odot}$ 
(see Fig.1b and Fig.1c), and the effect of very massive clusters
of $M>10^{15}M_{\odot}$ and low-mass halos below $10^{14}M_{\odot}$
is actually negligible. In other words, 
the SXRB, after AGNs are removed, is dominated by the  
baryons in the form of the WHIM and within groups and poor clusters. 
Of course, our present analytic models of the WHIM are still 
too simplistic, in the sense that a considerable fraction of
the WHIM may reside outside of less massive dark halos 
like groups and poor clusters due to nongravitational
processes such as preheating and/or radiative cooling. 
The inclusion of the WHIM outside the virial radii of dark halos  
in an analytic treatment is somewhat difficult because 
hydrostatic equilibrium may break down even if the NFW 
profile can be extrapolated to large radii.
One of the possibilities of extracting information about the WHIM 
residing outside of virial radii is to subtract the 
contribution of the WHIM within virial radii from the
observed SXRB in conjunction with theoretical modeling 
as we have done in this study.

Another approach to improving current work is to cross-correlate
the SXRB and the SZ sky, which may allow us to acquire additional 
knowledge of the distribution and evolution of the WHIM. Recall that
X-ray emission and SZ signals have very different dependence on
both the radial distribution of the WHIM in halos and 
the spatial locations of halos themselves.   
In fact, cross-correlations between the SXRB and other sources (galaxies,
clusters, CMB, radio sources, IRAS sources, etc.) have been 
extensively explored in literature. A combination of 
these cross-correlation and auto-correlation analyses of the SXRB 
may eventually constitute a powerful tool of unveiling the amount 
and distribution of the missing baryons in the local universe.

\acknowledgments
Useful discussion with Peng-Jie Zhang, Ue-Li Pen and Asantha Cooray
and valuable suggestions by an anonymous referee are gratefully 
acknowledged. We thank the hospitality of the Institute of Astronomy 
and Astrophysics, Academia Sinica, where part of this research was made. 
This work was supported by the National Science Foundation of China, 
under Grant No. 10233040, 
and the Ministry of Science and Technology of China, under Grant
No. NKBRSF G19990754.


\clearpage

\begin{figure}
\figurenum{1a}
\epsscale{0.8}   
\plotone{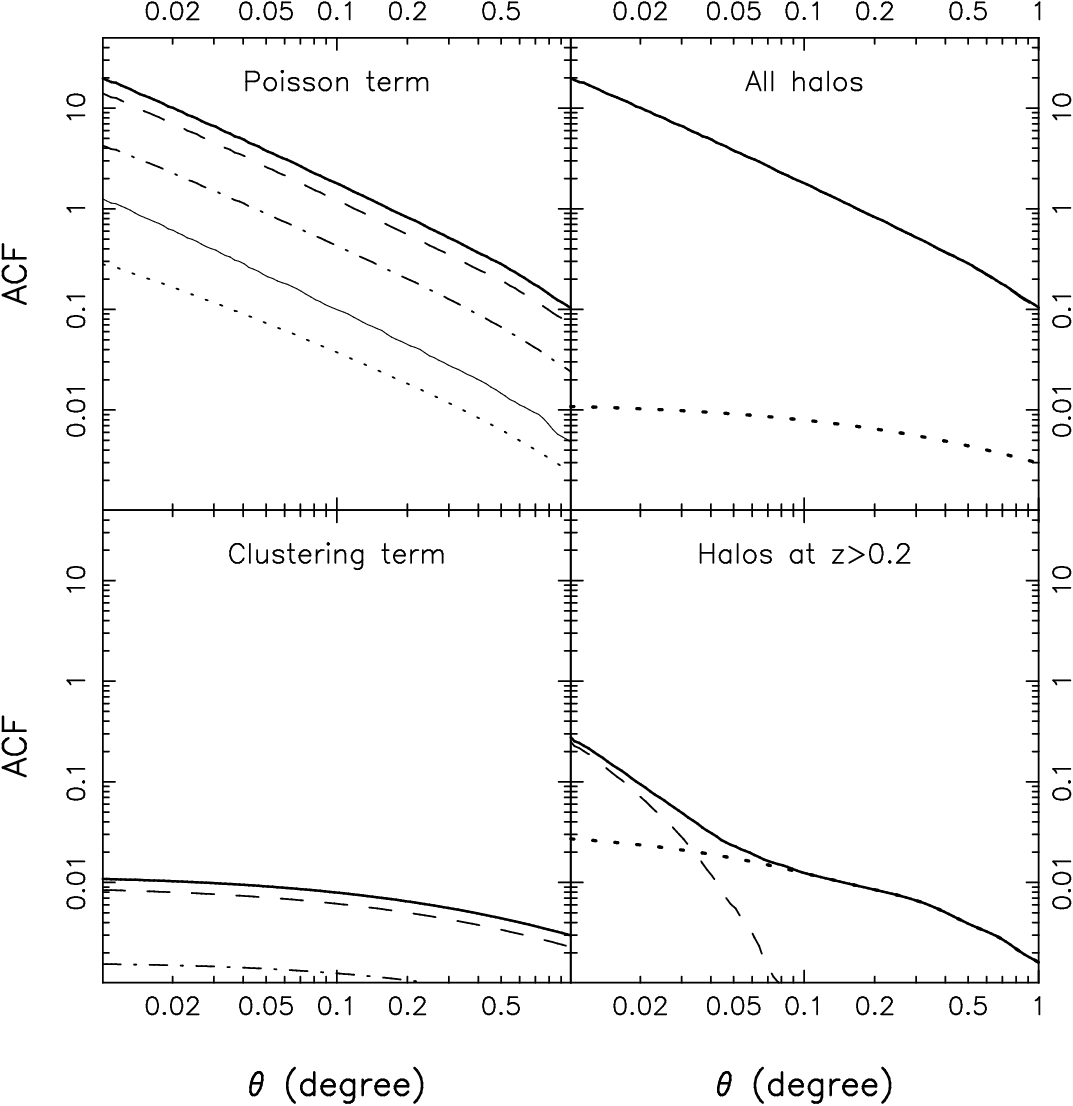}
\caption{Angular two-point autocorrelation function (ACF) of 
the soft X-ray background 
(SXRB) predicted by (a) self-similar model, (b) preheating model and
(c) cooling model.  Contributions of halos from different mass ranges 
to Poisson and clustering terms are demonstrated in left panels: 
thin solid line-($10^{12}-10^{13}M_{\odot}$); 
dashed line-($10^{13}-10^{14}M_{\odot}$);
dot-dashed line-($10^{14}-10^{15}M_{\odot}$);
and dotted line-($10^{15}-10^{16}M_{\odot}$).
ACFs from halos with and without exclusion of local population
within $z=0.2$ are displayed in
right panels, in which dashed and dotted lines represent 
the contributions of Poisson and clustering terms, respectively. 
\label{fig1}}
\end{figure}

\clearpage
\begin{figure}
\figurenum{1b}
\plotone{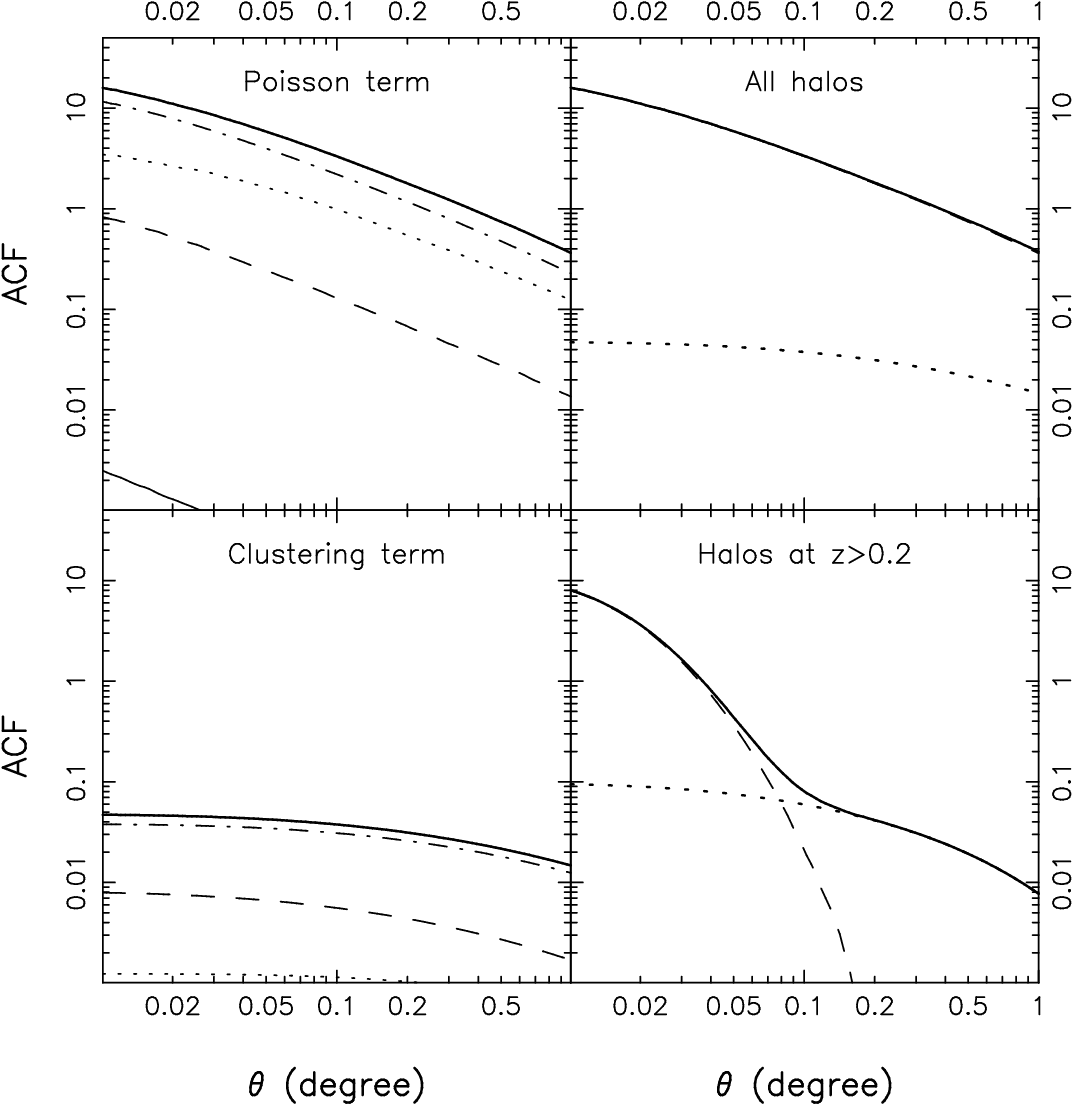}
\caption{ }
\end{figure}

\clearpage
\begin{figure*}
\figurenum{1c}
\plotone{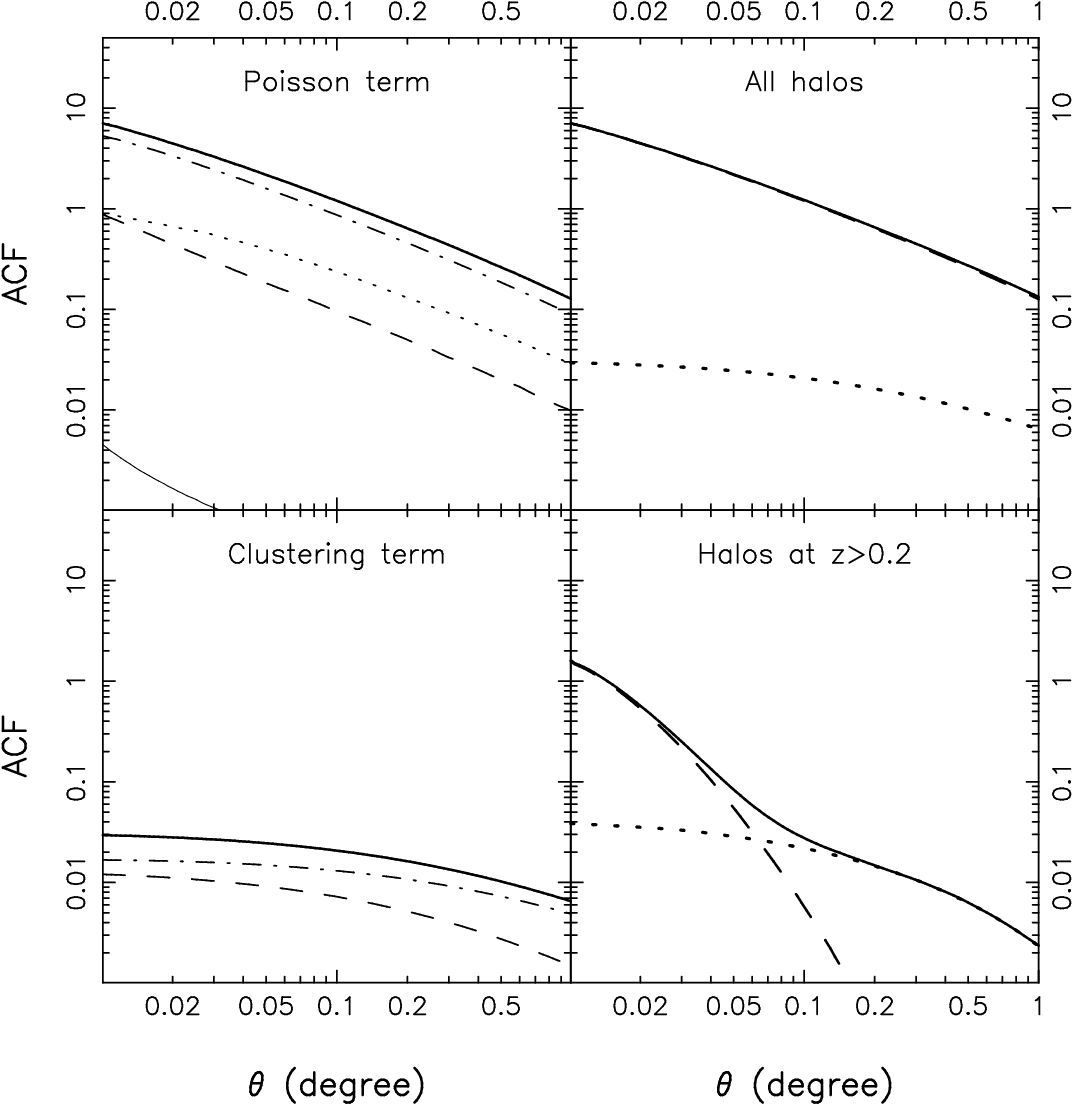}
\caption{ }
\end{figure*}

\clearpage
\begin{figure}
\figurenum{2}
\epsscale{0.60}    
\plotone{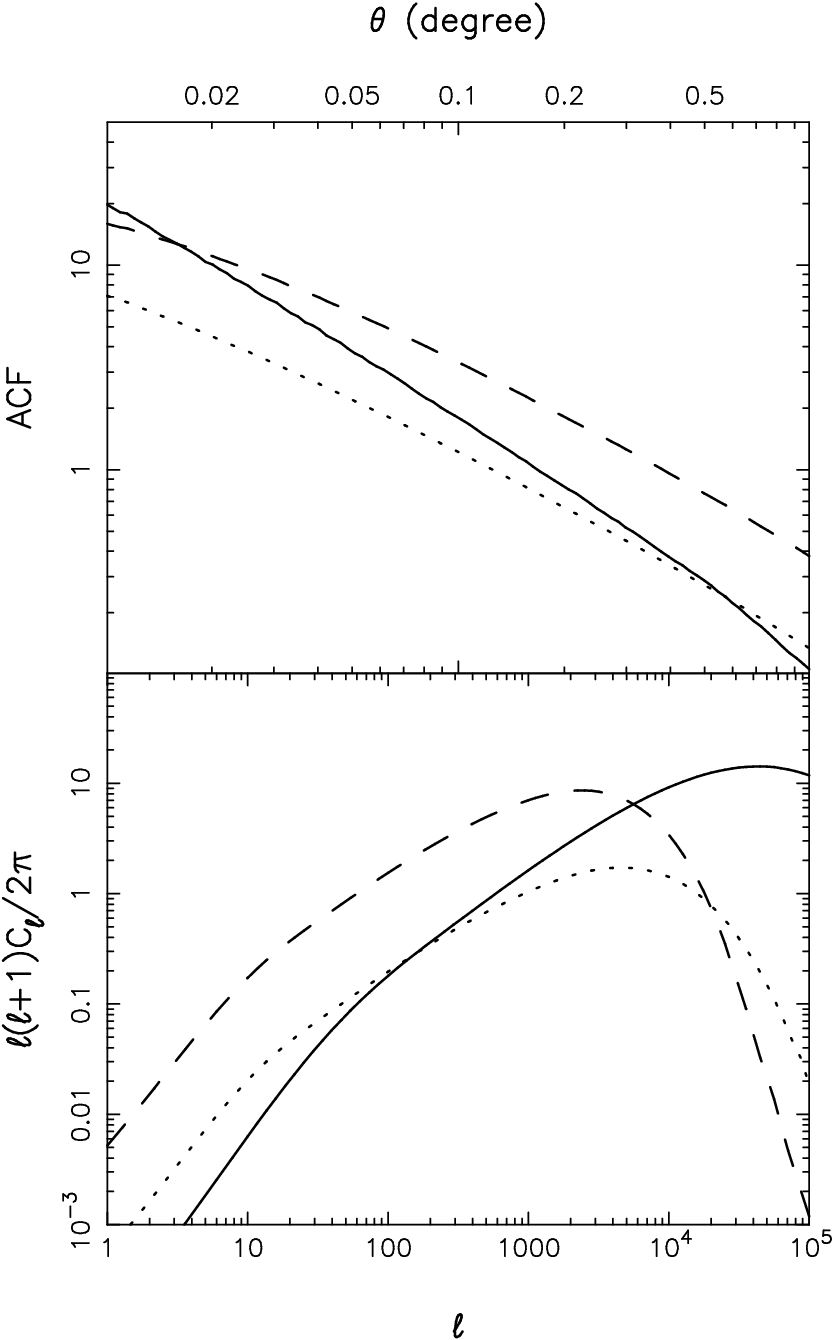}
\caption{Comparison of the ACFs (top panel) and power spectra
(lower panel) of the SXRB predicted by 
self-similar model (solid line), preheating model (dashed line)
and cooling model (dotted line). 
\label{fig2}}
\end{figure}

\clearpage
\begin{figure}
\figurenum{3}
\epsscale{0.55} 
\plotone{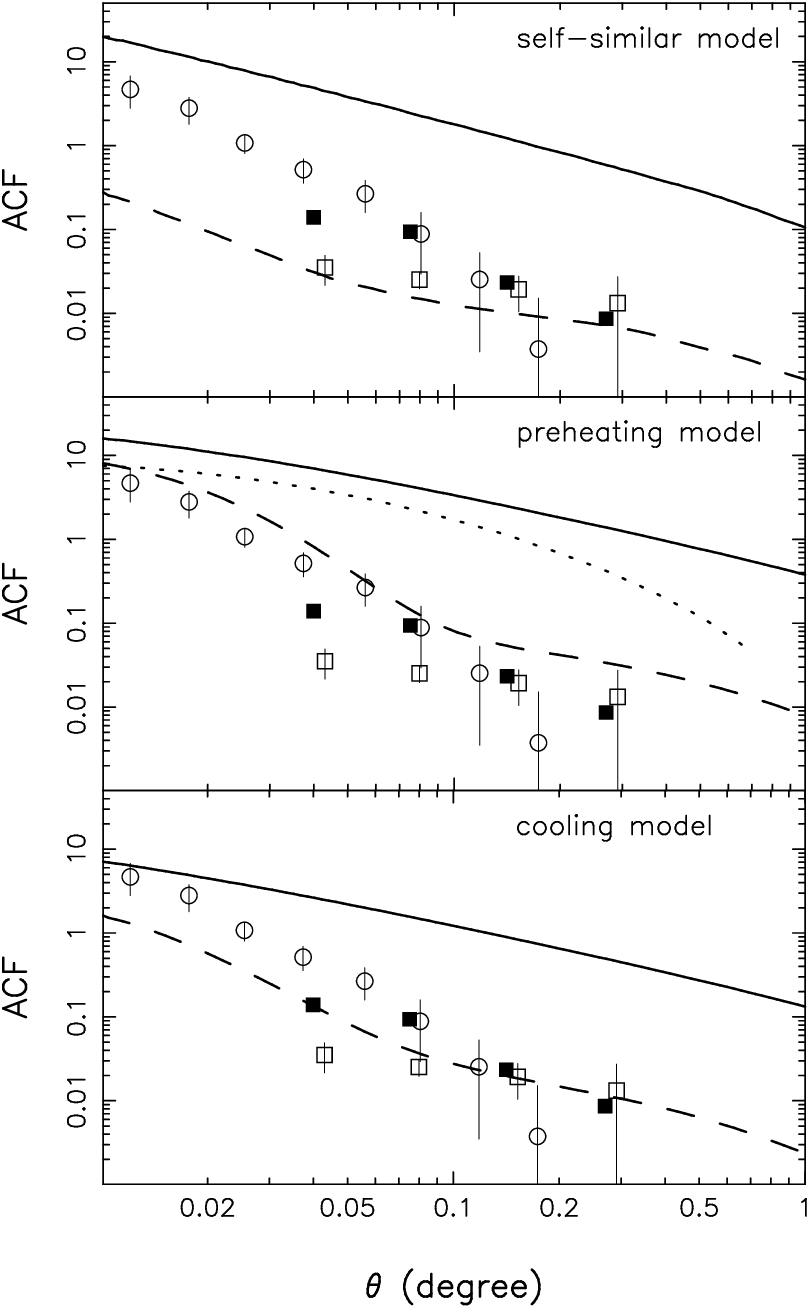}
\figcaption{The theoretically predicted ACFs of the SXRB are compared with 
the observationally determined data (Soltan et al. 2001) and the simulated
results (Croft et al. 2001; open circles). The model predicted ACFs 
with and without inclusion of nearby halos within $z=0.2$ 
are plotted by solid and dashed lines, respectively. 
Filled and open squares represent 
the data with and without inclusion of clusters in the 
fields of the ROSAT pointing observations used for the determination of
ACFs. Note that both observational and simulated data are dominated 
by AGNs rather than the WHIM. 
In the middle panel we have also illustrated the result of
Zhang \& Pen (2002) for preheating model with nongravitational
energy put $E=0.5$ keV per particle (dotted line).
\label{fig3}}
\end{figure}

\end{document}